\documentclass[epj, final, nopacs]{svjour}

\usepackage{graphicx}
\usepackage{amsmath}
\usepackage{amssymb}
\usepackage{xcolor}
\usepackage{microtype}
\usepackage{xspace}

\usepackage[T1]{fontenc} 
\usepackage{newtxtext, newtxmath} 

\usepackage{hyperref}
\usepackage{upgreek}
\usepackage{siunitx}
\usepackage{booktabs}
\usepackage{xurl} 
\hypersetup{colorlinks=true, allcolors=blue}

\usepackage{filecontents}

\makeatletter
\let\cl@chapter\undefined
\makeatother
\usepackage[capitalize]{cleveref}

\def\Offline{\mbox{$\overline{\textrm{Off}}$\hspace{.05em}\protect\raisebox{.4ex}{$\protect\underline{\textrm{line}}$}}\xspace}

\begin{document}

\title{A data-driven method for measuring corner-clipping probabilities in segmented particle detectors}

\author{
    J.~de~Jes\'us\inst{1,2,}\thanks{e-mail: joaquin.dejesus@usc.es (corresponding author)} \and
    J.~M.~Figueira\inst{1} \and
    F.~S\'anchez\inst{1} \and
    D.~Veberič\inst{2}
}

\institute{
    Instituto de Tecnolog\'ia en Detecci\'on y Astropart\'iculas (CNEA, CONICET, UNSAM), Buenos Aires, Argentina
    \and 
    Karlsruhe Institute of Technology, Institute for Astroparticle Physics, Karlsruhe, Germany
}

\date{Received: date / Revised version: date}

\abstract{The accuracy of particle counting in highly segmented detectors is limited by the corner-clipping effect, in which a single ionizing particle generates signals in adjacent detection elements. This phenomenon introduces a direction-dependent overcounting bias that distorts reconstructed observables and is commonly corrected using Monte-Carlo simulations, thereby inheriting modeling uncertainties. We present a fully data-driven method to directly measure the single-particle corner-clipping probability, exploiting the nanosecond timing resolution of modern detectors to statistically distinguish genuine corner-clipping events from random coincidences, with non-neighboring detection elements serving as an intrinsic control sample. The technique is validated using detailed simulations of the Underground Muon Detector of the Pierre Auger Observatory, reproducing the true angular dependence of the corner-clipping probability with absolute deviations below \num{0.01}. To parameterize the results, we introduce a compact analytical model incorporating detector geometry, minimum detectable path length, and orientation-independent contributions. The proposed methodology and parameterization enable the direct incorporation of data-driven corner-clipping corrections into reconstruction algorithms, mitigating the overcounting bias and ultimately yielding a more accurate determination of the muonic component of extensive air showers. These developments are broadly applicable to any segmented detector with sufficient timing resolution, making them relevant to a wide range of experiments in high-energy and astroparticle physics.}

\maketitle

\section{Introduction}
\label{sec:intro}

Large-area, highly segmented detectors are widely used in particle and astroparticle physics. The extensive arrays deployed in cosmic-ray observatories~\cite{Akeno:1992,CASA:1994,Akeno:1995,CASA:1999,AMIGA:2016} exemplify their indispensable capability to register and count incident particles over wide areas. In many such detectors, the primary observable is the number of activated detector segments, $k$, which serves as an experimental proxy for the true number of incident particles, $N$. The relationship between $k$ and $N$, however, is affected by several instrumental effects that can introduce systematic biases if not properly accounted for. These effects can be broadly categorized into two classes: undercounting and overcounting.

Undercounting arises primarily from the pile-up effect, in which two or more particles strike the same detector segment within the instrument's resolving time and are therefore recorded as a single hit. This effect is particularly pronounced in high-density particle environments. Because the statistical nature of pile-up is well understood, robust methodologies have been developed to mitigate this bias. These typically involve deriving the probability distribution of $k$ for a given expected number of incident particles and constructing an estimator for $N$ based on this distribution~\cite{Supanitsky:2008,Ravignani:2015,Supanitsky:2021,Gesualdi:2022,Varada:2023}.

A more subtle source of bias arises from an overcounting mechanism that we refer to as \textit{corner-clipping}, in which a single incident particle generates signals in multiple adjacent segments. This phenomenon can occur when an inclined particle traverses the boundary between two segments, or when a secondary particle, such as a knock-on electron produced by the primary particle~\cite{Scornavacche:2025m6}, deviates into a neighboring segment and deposits sufficient energy to activate it. Because the effect is intrinsically geometric, its probability increases with the zenith angle, $\theta$, of the incident particle and depends on the azimuthal angle, $\Delta\phi$, relative to the orientation of the detector segmentation. Uncorrected corner-clipping can lead to significant biases. For example, it may overestimate the muon content of air showers by $\qty{10}{\percent}$ or more at large zenith angles~\cite{PMT_paper}. Consequently, it introduces a complex, direction-dependent systematic effect that can distort physics measurements.

The standard approach to correcting corner-clipping relies on detailed Monte-Carlo simulations of the detector response~\cite{PMT_paper,Figueira:2017p}. Within this paradigm, the mean bias of an uncorrected estimator is parameterized as a function of the relevant geometric variables ($\theta$, $\Delta\phi$), and the resulting correction function is subsequently applied to experimental data. A key limitation of this approach is its dependence on the accuracy of the detector simulation. Simplifications or inaccuracies in the detector model, such as its geometry, material properties, or electronic response, propagate directly into the derived correction and ultimately into the physics results. Moreover, because the correction is typically obtained from the global detector response in simulation, it often combines corner-clipping with other effects, such as residual pile-up or segmentation biases, which hinders the isolation of the specific physical contribution.

In this work we introduce an alternative to simulation-based approaches for correcting corner-clipping effects. We present a novel, purely data-driven method that enables the direct measurement of the single-particle corner-clipping probability, $p_{\text{cc}}$. 
The method is applicable to any segmented detector whose timing resolution is significantly shorter than the characteristic inter-particle arrival time, allowing quasi-simultaneous signals produced by a single particle to be statistically distinguished from accidental coincidences. By isolating and quantifying this specific physical process using high-resolution timing information, the method decouples the corner-clipping correction from other systematic effects and obviates the reliance on detailed Monte-Carlo simulations.

To demonstrate the efficacy of the approach, we apply it to the Underground Muon Detector (UMD)~\cite{AMIGA:2016,PMT_paper,Scornavacche:2023,deJesus:2025} of the Pierre Auger Observatory~\cite{Auger:2015,Auger:2024,AMIGA:2025,AugerOpenData:2025}. 
Given its high segmentation and excellent timing performance, the UMD provides an ideal benchmark for validating the proposed method. Using UMD simulations as a testbed, we demonstrate a data-driven quantification of this effect, establishing a robust and broadly applicable methodology for segmented detectors.

The paper is organized as follows. In \cref{sec:method}, we present the data-driven technique used to estimate the corner-clipping probability and describe the statistical procedure employed to isolate corner-clipping events. In \cref{sec:model}, we introduce a physically motivated model that describes its angular dependence in terms of detector geometry and detection thresholds. In \cref{sec:application}, we apply the formalism to the UMD, using detailed Monte-Carlo simulations to validate both the methodology and the analytical parameterization, and to demonstrate an angle-resolved, data-driven determination of $p_{\text{cc}}$. Finally, in \cref{sec:conclusions}, we discuss the broader implications of this work and present our conclusions.

\section{Data-driven determination of the corner-clipping probability}
\label{sec:method}

In this Section we introduce a fully data-driven method for estimating the corner-clipping probability. The methodology assumes a segmented detector composed of contiguous elements arranged in a regular configuration and a timing resolution shorter than the typical inter-particle arrival time within the event. This requirement does not imply that corner-clipping coincidences can be uniquely identified on an event-by-event basis, since independent particles belonging to the same event may arrive with arbitrarily small time differences. Rather, the approach exploits the statistical structure of the time-difference distribution to identify regions in which signal correlations are negligible, allowing the contribution from independent particles to be determined directly from data. This statistical separation underpins the background estimation procedure introduced below.

\subsection{Detector segmentation and event topology}

\begin{figure}
\centering
\includegraphics[width=0.49\textwidth]{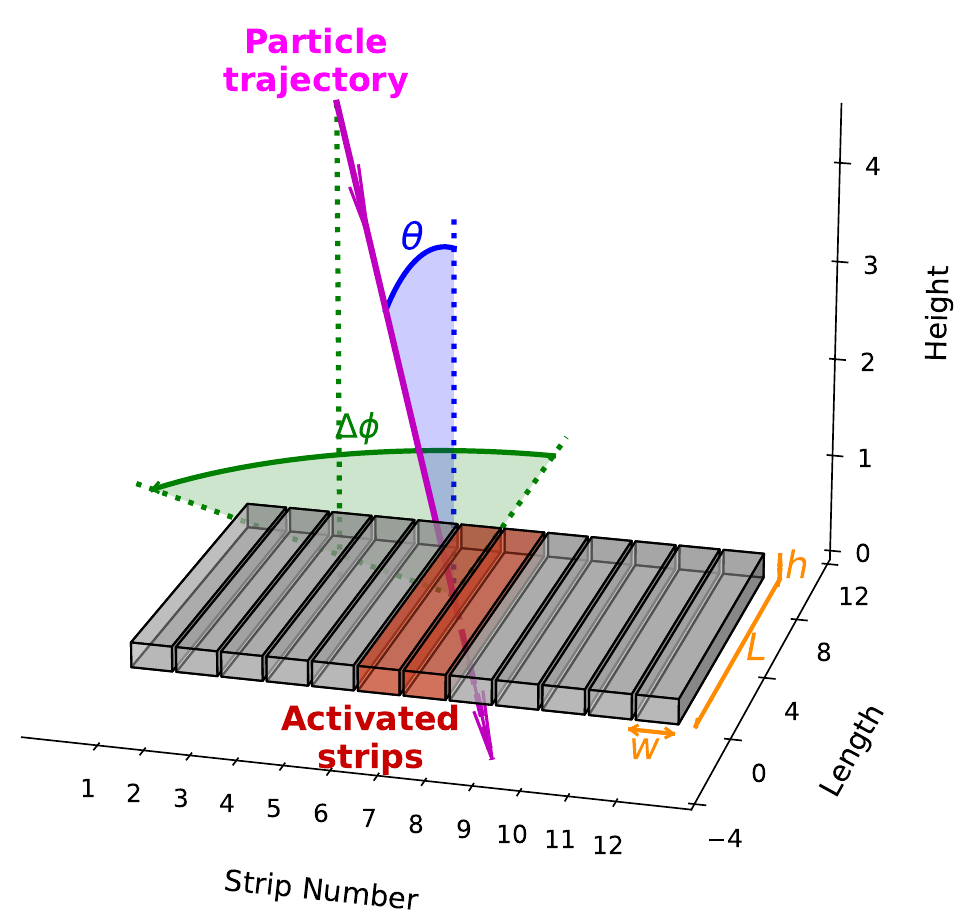}
\caption{Schematic of a generic particle detector composed of a linear array of contiguous segments. The trajectory of an incident particle is shown with a magenta arrow. Its direction is characterized by the zenith angle, $\theta$ (blue), and the azimuthal angle, $\Delta\phi$ (green), measured with respect to the longitudinal axis of the segments. The illustrated event corresponds to a corner-clipping case, in which the particle intersects two adjacent segments, producing simultaneous activations (highlighted in red).}
\label{fig:strip_detector}
\end{figure}

Consider a generic segmented particle detector in which each detection channel (segment) operates in counting mode. An illustrative case is shown in \cref{fig:strip_detector}, where the detector is represented as a linear array of contiguous segments. For the implementation of the methodology, the detector is organized into one or more discrete sub-units. This partitioning typically reflects the physical assembly of the module, which may consist of separate mechanical components that do not share segment boundaries. Since corner-clipping is inherently a local phenomenon requiring the activation of adjacent elements, segments belonging to different mechanical units are treated as independent sampling domains for the estimator.

To distinguish between the different hierarchical scales of the detector, variables associated with the sub-units are denoted with a tilde ($\tilde{\cdot}$), while whole-module quantities remain unadorned. Let each sub-unit consist of $\tilde{n}_{\text{s}}$ segments. For a given event, the relevant observable is $\tilde{k}$, the total number of activated segments within a single sub-unit. The method focuses on events where exactly two segments are activated ($\tilde{k}=2$, hereafter referred to as two-segment events). These events are classified into two mutually exclusive and topologically distinct categories: \emph{neighboring pairs}, in which the activated segments are physically adjacent, and \emph{non-neighboring pairs}, in which they are separated by one or more inactive segments.

\subsection{Time structure of signal and background}
The central premise of the method is that the distribution of absolute time differences, $\lvert \Delta t \rvert$, between signals in two-segment events exhibits distinct characteristics for neighboring and non-neighboring pairs. This difference originates from the underlying physical processes responsible for the activations.

For non-neighboring pairs, both segments are activated by independent particles from the same primary event (e.g., distinct muons from an air shower). Their $\lvert \Delta t \rvert$ distribution therefore reflects the intrinsic temporal spread of the shower front.

Neighboring pairs, however, consist of two distinct contributions. One originates from two independent particles that happen to strike adjacent segments. The other arises from genuine corner-clipping events, in which a single particle simultaneously activates two adjacent segments. The temporal characteristics of the two populations are distinct: corner-clipping events yield signals that are nearly simultaneous, producing a narrow peak at $\lvert \Delta t \rvert \approx 0$ whose width is governed by the instrumental time resolution (scintillator response, light propagation, and electronics). In contrast, independent-particle activations yield the same broad $\lvert \Delta t \rvert$ distribution observed in non-neighboring pairs.

Consequently, a statistically significant excess in the fraction of neighboring pairs at small $\lvert\Delta t\rvert$ constitutes a robust signature of correlated neighboring activations and enables the corner-clipping contribution to be estimated.

\subsection{Data-driven estimation of the corner-clipping contribution}
The quantification of corner-clipping events proceeds as follows. We first consider all sub-units in which exactly two segments are activated ($\tilde{k}=2$). The total number of such cases is denoted by $N_{2\text{-segs}}$, which decomposes into neighboring and non-neighboring contributions,
\begin{equation}
N_{2\text{-segs}} = N_{\text{neigh}} + N_{\text{non-neigh}},
\label{eq:n2bars}
\end{equation}
where $N_{\text{neigh}}$ and $N_{\text{non-neigh}}$ represent the counts of neighboring and non-neighboring pairs, respectively. The neighboring sample consists of both genuine corner-clipping events and background from independent particles, i.e.
\begin{equation}
N_{\text{neigh}} = N_{\text{cc}} + N_{2\upmu},
\label{eq:nneigh_decomp}
\end{equation}
where $N_{\text{cc}}$ denotes the number of corner-clipping events and $N_{2\upmu}$ the number of neighboring activations arising from two independent penetrating particles. We adopt the subscript $\upmu$ to anticipate the specific application to muon detection in \cref{sec:application}, noting that the formalism itself remains valid for any type of charged particle. The central analytical task is therefore to disentangle the corner-clipping signal, $N_{\text{cc}}$, from the background contributions using directly observable quantities.

The key to separating the signal from the background is the probability $p_{2\upmu}$ that a two-particle event forms a neighboring pair. This probability can be obtained (\emph{i}) theoretically, from simple combinatorial arguments, or (\emph{ii}) empirically, from data in a control region free of corner-clipping contributions. From combinatorics, if two distinct segments are activated in a sub-unit of size $\tilde{n}_\text{s}$, there are $\binom{\tilde{n}_\text{s}}{2} = \tilde{n}_\text{s}(\tilde{n}_{\text{s}}-1)/2$ possible pairs, of which $\tilde{n}_{\text{s}}-1$ correspond to neighboring segments. The expected probability is thus
\begin{equation}
p_{2\upmu}^{\text{exp}} = \frac{\tilde{n}_{\text{s}}-1}{\binom{\tilde{n}_\text{s}}{2}} = \frac{2}{\tilde{n}_\text{s}}.
\label{eq:p2mu_exp}
\end{equation}
A fully data-driven estimate can be obtained by introducing the observed neighboring-pair fraction,
\begin{equation}
    p_{\text{neigh}}^{\text{obs}} = \frac{N_{\text{neigh}}}{N_{2\text{-segs}}},
    \label{eq:p_neigh}
\end{equation}
which quantifies the relative occurrence of neighboring activations among all two-segment events. As the time separation $\lvert \Delta t \rvert$ between activations increases, the contribution from corner-clipping events to $N_{\text{neigh}}$ diminishes and eventually becomes negligible ($N_{\text{cc}} \approx 0$). Consequently, $p_{\text{neigh}}^{\text{obs}}$ decreases until it stabilizes at a constant, asymptotic value corresponding to the pure two-particle background. This behavior defines a control region at large time differences, $\lvert \Delta t \rvert > \lvert \Delta t \rvert_{\text{cut}}$, with the threshold $\lvert \Delta t \rvert_{\text{cut}}$ identified as the point where $p_{\text{neigh}}^{\text{obs}}$ reaches its plateau.

Within this control region, where the contribution from corner-clipping is negligible, $p_{\text{neigh}}^{\text{obs}}$ directly estimates the background probability for two independent particles to form a neighboring pair. We therefore define the observed background probability as
\begin{equation}
    p_{2\upmu}^{\text{obs}} = p_{\text{neigh}}^{\text{obs}}, 
    \quad \text{for } \lvert \Delta t \rvert > \lvert \Delta t \rvert_{\text{cut}}.
    \label{eq:p2mu_meas_final}
\end{equation}
Agreement between this observed value and the combinatorial expectation $p_{2\upmu}^{\text{exp}}$ offers a stringent self-consistency check of the method.

In the signal region, $\lvert \Delta t \rvert < \lvert \Delta t \rvert_{\text{cut}}$, corner-clipping contributions become significant. In this regime, the number of two-segment events attributable to independent particles is $N_{2\text{-segs}} - N_{\text{cc}}$, while the number of neighboring pairs arising from independent particles is $N_{\text{neigh}} - N_{\text{cc}}$, as follows from \cref{eq:nneigh_decomp}. Consequently, the probability that two independent particles form a neighboring pair can be written as
\begin{equation}
p_{2\upmu}^{\text{obs}} = \frac{N_{\text{neigh}} - N_{\text{cc}}}{N_{2\text{-segs}} - N_{\text{cc}}}, 
\quad \text{for } \lvert \Delta t \rvert < \lvert \Delta t \rvert_{\text{cut}}.
\label{eq:p2mu_def2}
\end{equation}
Inverting this expression yields the direct estimator for the number of corner-clipping events
\begin{equation}
\hat{N}_{\text{cc}} = \frac{N_{\text{neigh}} - p_{2\upmu}^{\text{obs}} \, N_{2\text{-segs}}}{1 - p_{2\upmu}^{\text{obs}}}.
\label{eq:Ncc_final2}
\end{equation}
The corresponding number of neighboring pairs attributable to two independent particles then follows as
\begin{equation}
\hat{N}_{2\upmu} = N_{\text{neigh}} - \hat{N}_{\text{cc}}.
\label{eq:N2mu_final2}
\end{equation}
Thus, these estimators depend only on measurable quantities: the number of neighboring pairs $N_{\text{neigh}}$, the total number of two-segment events $N_{2\text{-segs}}$, and the background probability $p_{2\upmu}^{\text{obs}}$, obtained either from the control region or from \cref{eq:p2mu_exp}.

The formalism can be applied on a $\lvert \Delta t \rvert$ bin-by-bin basis, allowing the neighboring and non-neighboring pairs, the neighboring-pair fraction, and the corner-clipping contribution to be resolved as functions of time separation. For each bin $\lvert \Delta t \rvert_i$, the observables are determined, the estimator $\hat{N}_{\text{cc}}(\lvert \Delta t \rvert_i)$ evaluated, and the corresponding contribution extracted. Summing the individual contributions over the signal region yields the total number of corner-clipping events
\begin{equation}
\hat{N}_{\text{cc}}^{\text{tot}} = \sum_{\lvert \Delta t \rvert_i \leq \lvert \Delta t \rvert_{\text{cut}}} \hat{N}_{\text{cc}}(\lvert \Delta t \rvert_i).
\end{equation}
In this way, the corner-clipping contribution can be quantified directly from experimental data while preserving the temporal information encoded in the $\lvert \Delta t \rvert$ distribution.

\subsection{Definition and estimation of the corner-clipping probability}
Once the total number of corner-clipping events has been estimated, we introduce the physically relevant quantity: the single-particle corner-clipping probability, $p_{\text{cc}}$. This probability is defined as the likelihood that a single particle incident on a sub-unit produces a corner-clipping signature, i.e. $\tilde{k}=2$ with neighboring activations. If $N_{1\upmu}$ denotes the total number of single-particle incidences on a sub-unit, then
\begin{equation}
p_{\text{cc}} = \frac{N_{\text{cc}}^{\text{tot}}}{N_{1\upmu}}.
\end{equation}

To evaluate $p_{\text{cc}}$ from data, an estimate of $N_{1\upmu}$ is required. Under the assumptions of negligible electronic noise and detection inefficiency, the vast majority of single-particle events result either in the activation of a single segment or in a corner-clipping event. Thus, one can approximate
\begin{equation}
N_{1\upmu} \approx N_{1\text{-seg}} + N_{\text{cc}}^{\text{tot}},
\end{equation}
where $N_{1\text{-seg}}$ denotes the number of sub-units with exactly one activated segment (i.e. $\tilde{k}=1$), a quantity directly measurable from data.

This leads to the fully data-driven estimator for the corner-clipping probability
\begin{equation}
\hat{p}_{\text{cc}} = \frac{\hat{N}_{\text{cc}}^{\text{tot}}}{N_{1\text{-seg}} + \hat{N}_{\text{cc}}^{\text{tot}}}.
\label{eq:pcc_est}
\end{equation}
This estimator can be evaluated as a function of the incident particle direction to quantify the bias introduced by corner-clipping. In this way, the method enables a direct, fully data-driven determination of the corner-clipping probability, $p_{\text{cc}}$, from experimental data, without reliance on detector simulations.

While this method provides an empirical characterization of the effect, its interpretation requires a model that relates the observed angular dependence to detector geometry and detection thresholds. This is addressed in the following section.

\section{Corner-clipping probability model}
\label{sec:model}

In this section we develop a model for the corner-clipping probability, aimed at explaining its directional dependence and relating it to the detector and segment geometry, as well as to the minimum path length required for signal detection. The derivation proceeds in two stages: first for an idealized detector geometry, in which any particle trajectory intersecting two adjacent segments produces a signal, and subsequently for a realistic detector, where finite path-length thresholds and secondary-particle contributions must be taken into account.

We consider a detector segmented into contiguous rectangular scintillator strips of length $L$, width $w$, and height $h$. The direction of an incident particle is characterized by a zenith angle $\theta$, measured relative to the normal of the detector plane, and an azimuthal angle $\Delta\phi$, defined with respect to the longitudinal axis of the segment, as illustrated in \cref{fig:strip_detector}. In the following, the corner-clipping probability is therefore expressed as $p_{\text{cc}}(\theta,\Delta\phi)$.

We begin by isolating the purely geometric contribution to the corner-clipping probability, neglecting detector response effects. This idealized treatment provides a reference against which the impact of finite detection thresholds and additional physical processes can be assessed.

\subsection{Idealized detector}
In the idealized approximation, a corner-clipping event occurs whenever a particle trajectory intersects both the top surface of a strip and the lateral boundary shared with its neighbor, irrespective of the path length traversed. The corner-clipping probability is then defined as the ratio between the rate of particles crossing the lateral boundary, $R_{\text{lat}}$, and the rate crossing the top surface, $R_{\text{top}}$,
\begin{equation}
p_{\text{cc}}(\theta,\Delta\phi)
= \frac{R_{\text{lat}}}{R_{\text{top}}}.
\end{equation}
Assuming a uniform incident flux, these rates are proportional to the corresponding effective areas, defined as the geometric areas of the surfaces projected onto a plane perpendicular to the particle momentum. The probability can therefore be written as
\begin{equation}
p_{\text{cc}}(\theta,\Delta\phi)
= \frac{A_{\text{lat}}^{\text{eff}}}{A_{\text{top}}^{\text{eff}}},
\end{equation}
where $A_{\text{lat}}^{\text{eff}}$ and $A_{\text{top}}^{\text{eff}}$ refer to the effective areas of the lateral boundary and top surface, respectively. 

Each effective area is determined by the orientation of the surface normal relative to the particle momentum, with the detector geometry and the definitions of the angles $\theta$ and $\Delta\phi$ illustrated in \cref{fig:strip_detector}. 
The top surface, with area $wL$, has a surface normal perpendicular to the detector plane, yielding a projection factor of $\cos\theta$ and the effective area 
\begin{equation} A_{\text{top}}^{\text{eff}} = wL\cos\theta. 
\label{eq:A_top_eff} 
\end{equation} 
The lateral boundary, with area $hL$, has a surface normal that lies in the detector plane and is perpendicular to the longitudinal axis of the strip. The projection of this surface onto a plane normal to the particle momentum depends on both the zenith and azimuthal angles, yielding the effective area 
\begin{equation} A_{\text{lat}}^{\text{eff}} = hL\sin\theta\sin\Delta\phi. 
\label{eq:A_lat_eff} 
\end{equation}
The ideal corner-clipping probability thus becomes
\begin{equation}
p_{\text{cc}}^{\text{ideal}}(\theta,\Delta\phi)
= \frac{h}{w}\tan\theta\,\sin\Delta\phi.
\end{equation}

This expression represents the purely geometric baseline for the corner-clipping probability. In a real detector, however, additional effects modify this behavior. Finite detection thresholds reduce the probability by suppressing trajectories with insufficient path length, while secondary particles generated along the primary track can enhance the probability by producing additional correlated activations. The net corner-clipping probability therefore reflects the interplay of these competing effects, which are addressed in the following subsections.

\begin{figure}
    \centering
    \includegraphics[width=0.495\textwidth]{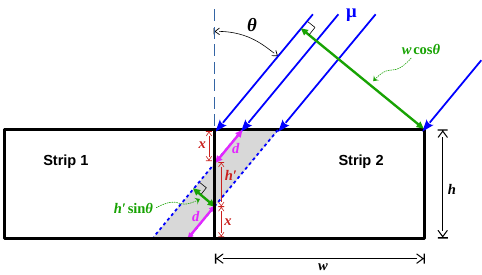}
    \caption{Schematic side view (cross-section) of two adjacent detector strips of height $h$ and width $w$, illustrating the geometric model for corner-clipping. A corner-clipping event is recorded only when an incident particle ($\upmu$) traverses a minimum path length $d$ in both strips. For a given zenith angle $\theta$, this requirement defines excluded regions of height $x$ at the upper and lower edges of the shared boundary, leaving a central active region of height $h'$ (shaded). The projection of this active region onto a plane perpendicular to the particle direction is $h'\sin\theta$, while the projection of the top surface onto the same plane is $w\cos\theta$.}
    \label{fig:model_schematic}
\end{figure}

\subsection{Real detector and finite path-length effects}
In a real detector, a segment is triggered only if the particle deposits sufficient energy, corresponding to a minimum path length $d$ within the scintillator. For a corner-clipping event to be registered, this condition must be met in both adjacent strips. It should be noted that the parameter $d$ represents an effective average over the strip, as light attenuation causes the actual minimum path length required for triggering to depend on the impact position of the particle relative to the photodetector.
As shown in \cref{fig:model_schematic}, this requirement creates excluded regions near the top and bottom of the shared boundary, where the particle path length in one of the segments falls below threshold. For a given $\theta$, the vertical extent of each excluded region is given by the projection of the minimum path length onto the detector’s vertical axis, 
$x = d\cos\theta$.
The active vertical region of the boundary is therefore reduced from $h$ to
$h' = h - 2x = h - 2d\cos\theta$.
Only particles intersecting the boundary within this central region contribute to detectable corner-clipping events. The corresponding effective lateral area for a real detector becomes
\begin{equation}
A_{\text{lat}}^{\text{eff}} = h' \sin\theta \, \, L  \sin\Delta\phi = (h - 2d\cos\theta) \sin\theta \, \, L \sin\Delta\phi,
\end{equation}
while the effective top area remains $A_{\text{top}}^{\text{eff}} = wL\cos\theta$. 

The resulting expression for the corner-clipping probability is thus
\begin{equation}
p_{\text{cc}}^{\text{real}}(\theta,\Delta\phi)
= \frac{A_{\text{lat}}^{\text{eff}}}{A_{\text{top}}^{\text{eff}}}
= \left(\frac{h}{w}\tan\theta - \frac{2d}{w}\sin\theta\right)\sin\Delta\phi.
\end{equation}
The second term in parentheses represents the efficiency loss near the segment edges due to the finite detection threshold. In the limit $d \to 0$, this correction vanishes and the expression reduces to the ideal geometric case.

\subsection{Contribution from secondary particles}
The model derived thus far accounts strictly for the primary particle trajectory. In practice, however, a competing effect arises from secondary particles (such as knock-on electrons) produced by the primary particle. These secondaries can trigger signals in neighboring segments even when the primary trajectory does not intersect the shared boundary, contributing an additional component to the measured neighboring-pair probability. To a first approximation, this contribution is represented by an orientation-independent additive term, $p_{\text{cc}}^{\text{sec}}$. Including this term, the complete expression for the corner-clipping probability becomes
\begin{equation}
p_{\text{cc}}^{\text{real}}(\theta,\Delta\phi)
= \left(\frac{h}{w}\tan\theta - \frac{2d}{w}\sin\theta\right)\sin\Delta\phi + p_{\text{cc}}^{\text{sec}}.
\label{eq:pcc_param1}
\end{equation}
This formulation explicitly expresses the dependence of $p_{\text{cc}}$ on the segment geometry ($h$, $w$), the minimum detectable path length ($d$), and the orientation-independent contribution from secondary interactions ($p_{\text{cc}}^{\text{sec}}$). It provides a direct physical interpretation of the measured corner-clipping probability and a practical basis for implementing reconstruction corrections in segmented detectors. For $\theta \lesssim \ang{30}$, the functions $\tan\theta$ and $\sin\theta$ are numerically similar (their ratio $\sec\theta$ departs from unity by less than $\qty{15}{\percent}$) rendering the two terms in \cref{eq:pcc_param1} nearly degenerate in that regime. While this similarity does not affect the model's ability to describe the total corner-clipping probability, it implies that the terms are not strictly orthogonal at low angles. Consequently, when fitting the model to data, a high correlation between the coefficients $A = h/w$ and $B = -2d/w$ is expected, reflecting a partial degeneracy in how the geometric and threshold-induced effects are statistically disentangled. Accurately isolating these contributions therefore requires measurements at sufficiently large $\theta$, where the two trigonometric functions diverge. This point is revisited quantitatively in \cref{sec:application}.

The performance of the data-driven estimator introduced in the previous section and the descriptive capability of this model are examined in the following section.

\section{Simulation-based validation with the Underground Muon Detector}
\label{sec:application}

The proposed methodology is applied to simulated data from the Underground Muon Detector (UMD) of the Pierre Auger Observatory, whose segmentation and timing resolution make it particularly well suited for validation. Using a detailed simulation chain, we characterize the corner-clipping effect and we evaluate the performance of both the data-driven estimator and the analytical model. While the formalism applies to any penetrating charged particle, the UMD is designed to detect muons, and the results presented below therefore refer exclusively to muons as the incident particles.

\subsection{Detector description}
The UMD is composed of buried plastic scintillation counters, each paired with a water-Cherenkov surface detector (SD) station of the Pierre Auger Observatory. These hybrid detector stations are deployed on a \qty{750}{\m} triangular grid covering an area of \qty{23.5}{\km\squared}, which is fully efficient for air showers with energies above \qty{3e17}{\eV} and zenith angles $\theta \le \ang{55}$. A denser \qty{433}{\m} grid, nested within this array, extends over \qty{1.9}{\km\squared} and lowers the detection threshold to \qty{6e16}{\eV}. 

The scintillator planes are buried at a depth of \qty{2.3}{\m} beneath the calcareous sedimentary soil of the Pampa Amarilla site, which has a density of \qty{2.35}{\g\per\cm\cubed} and a radiation length of \qty{24.5}{\g\per\cm\squared}. This results in a vertical overburden of approximately \qty{540}{\g\per\cm\squared}, which is equivalent to \num{22} radiation lengths. This thickness suppresses the electromagnetic component to a negligible level for the purposes of the UMD muon measurement. Furthermore, as this thickness represents approximately \num{5} nuclear interaction lengths, the soil significantly attenuates the hadronic component, ensuring that essentially only muons (along with a negligible fraction of high-energy hadrons) reach the buried detectors. Consequently, the shielding imposes an energy threshold of about \qty{1}{\GeV} for vertical muons.

\begin{figure*}
	\centering
	\includegraphics[width=.95\textwidth]{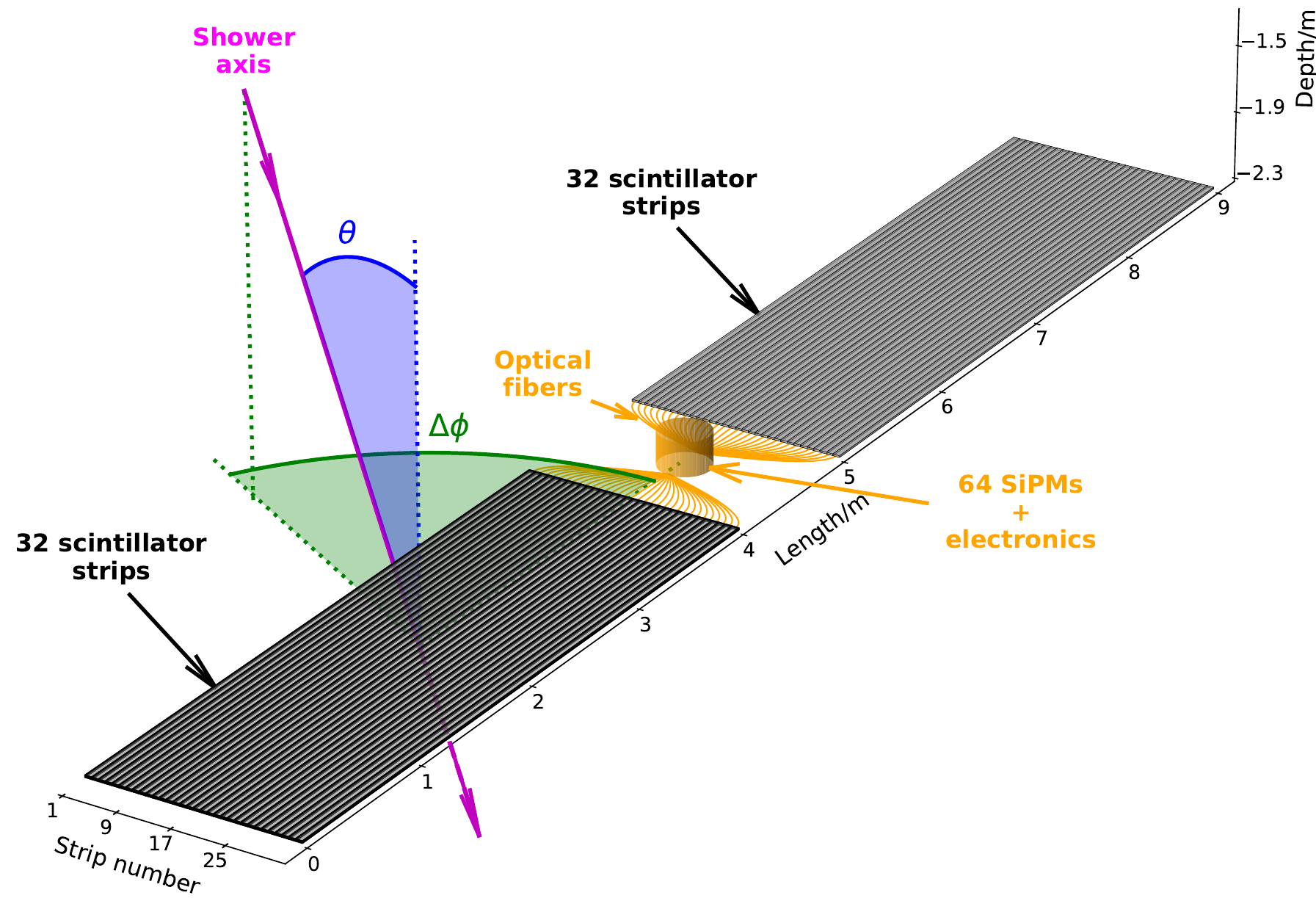}
	\caption{Schematic of an Underground Muon Detector module. The module is segmented into 64 scintillator strips and is mechanically divided into two 32-strip sub-units. The incident particle trajectory is approximated by the shower axis (magenta), which is defined by the zenith angle, $\theta$ (blue), and the azimuthal angle, $\Delta\phi$ (green), relative to the longitudinal axis of the strips.}
	\label{fig:UMD_Schematics}
\end{figure*}

Each UMD station consists of three \qty{10}{\meter\squared} scintillator modules. As shown schematically in \cref{fig:UMD_Schematics}, a standard module is segmented into 64 plastic scintillator strips (\qtyproduct{400 x 4.1 x 1.0}{\cm}). Scintillation light produced by a traversing muon is captured by embedded wavelength-shifting fibers and transported to silicon photomultipliers (SiPMs).

The UMD implements two complementary acquisition modes. The primary mode for this work is the binary (counter) mode, which leverages the detector's fine segmentation by processing each of the 64 SiPM channels independently. In this mode, signals are pre-amplified, shaped, and discriminated. Upon receiving a local station trigger from the companion SD station, a Field-Programmable Gate Array (FPGA) digitizes the discriminator outputs at \qty{320}{\MHz} (\qty{3.125}{\nano\second} sampling interval), yielding a \qty{6.4}{\micro\second} binary trace for each channel. A sample value of 1 indicates the signal exceeded the threshold. Offline, a muon hit is identified by a \emph{single-muon pattern}, defined as four or more consecutive 1s~\cite{AMIGA:2021}. A complementary ADC (calorimetric) mode~\cite{ADC:2025}, optimized for high muon densities, treats each module as an integrated unit by summing the signals from all SiPMs. The aggregated signal is amplified and subsequently digitized at \qty{6.25}{\nano\second} intervals, generating waveforms comprising 1024 samples each. The muon count is reconstructed by dividing the integrated charge of the waveform by the mean charge deposited by a single muon. 

This dual-mode architecture combines fine-grained counting capability in the binary mode with a large dynamic range in the ADC mode. The combination of high segmentation, counter-mode operation, and nanosecond-level timing resolution makes the UMD a particularly well-suited instrument for precision measurements of the muonic component of extensive air showers, and an ideal benchmark for the methodology developed in this work. Realizing the full potential of the binary mode, however, requires mitigating the overcounting introduced by the corner-clipping effect. The data-driven, timing-based procedure proposed here is specifically designed to isolate and correct this instrumental systematic, thereby establishing the binary mode as a high-fidelity muon counter.

In many large-area cosmic-ray experiments, the directions of individual muons within an extensive air shower cannot be measured directly. Instead, the shower axis is reconstructed from the relative timing of the surface detector stations, providing the incident direction of the primary cosmic ray. The muon-related observables are then analyzed as functions of this axis orientation. Simulations show that the angular spread of muons around the shower axis is typically limited to a few degrees, such that the average muon propagation direction closely follows the reconstructed axis.
Consequently, instrumental effects that depend on particle direction, such as the overcounting bias from corner-clipping, are primarily governed by the orientation of the shower axis. The corner-clipping probability, $p_{\text{cc}}$, must therefore be expressed as a function of this quantity. This situation applies directly to the UMD, where individual muon trajectories cannot be reconstructed. Accordingly, $p_{\text{cc}}$ is quantified as a function of the shower axis orientation relative to the module, defined by the zenith angle, $\theta$, and the azimuthal angle, $\Delta\phi$. These reference angles, illustrated for a typical corner-clipping event in \cref{fig:UMD_Schematics}, are reconstructed for each event by the surface detector of the Pierre Auger Observatory~\cite{Auger:2020sd,SD:2021,SSD:2025}.

An additional feature of the UMD module design, illustrated in \cref{fig:UMD_Schematics}, is its division into two mechanically independent halves, each containing 32 strips. These halves naturally define the sub-units treated as independent sampling domains for the estimator, such that $\tilde{n}_\text{s} = 32$. For this configuration, the expected probability that two independent muons striking random strips will activate a neighboring pair, according to \cref{eq:p2mu_exp}, is $p_{2\upmu}^{\text{exp}} = 2/32 = \qty{6.25}{\percent}$.

Although the present study focuses on the UMD and muons from extensive air showers, the analysis framework is not specific to this detector. Its application relies only on segmentation, a well-defined adjacency between detection elements, and timing resolution sufficient to statistically distinguish correlated activations from accidental coincidences, making the methodology applicable to a wide class of segmented detectors.

\subsection{Monte-Carlo validation of the method}
\label{subsec:mc_validation}
The performance of the method was validated using a detailed Monte-Carlo (MC) simulation chain designed to reproduce both the physics of extensive air showers and the detector response of the UMD. A key advantage of this approach is that the ground truth is explicitly known: the number of muons impinging on each detector segment, together with their arrival times and trajectories, is determined exactly. This enables a direct evaluation of the method's ability to identify and quantify corner-clipping events under controlled conditions, allowing us to test whether the estimator recovers the expected corner-clipping probability, $p_{\text{cc}}$, and whether it remains robust against variations in shower geometry and primary energy. Such a validation is essential before applying the method to experimental data, where the true underlying particle information is not directly accessible.

To carry out this validation, we employed a library of simulated showers generated with the \textsc{Corsika} framework~\cite{Heck:1998}, using \textsc{Epos-LHC}~\cite{Pierog:2015} as the high-energy hadronic interaction model. The library includes proton and iron primaries at three discrete energies, \qtylist[list-units=single, parse-numbers=false]{10^{17.5}; 10^{18}; 10^{18.5}}{\eV}, and seven zenith angles, \qtylist{0; 12; 22; 32; 38; 48; 56}{\degree}. For each combination of primary, energy, and zenith angle, 120 showers were available, resulting in a total of 5040 simulated events. The azimuthal angle of each shower was uniformly distributed over the full range, between $\ang{0}$ and $\ang{360}$. The simulated phase space was chosen to match and extend the operational range of the UMD. The lowest simulated energy, \qty[parse-numbers=false]{10^{17.5}}{eV}, lies just above the full-efficiency threshold of the \qty{750}{\meter} surface detector array. The zenith angle coverage extends up to \ang{56}, deliberately exceeding the $\theta \approx \ang{45}$ limit originally envisaged for UMD operation \cite{PMT_paper}.
This extended range allows us to test whether the data-driven methodology developed in this work can reliably address the corner-clipping bias at large zenith angles, where the effect is most pronounced.

The secondary particles from the \textsc{Corsika} showers were propagated through a full detector model implemented in \textsc{Geant4} \cite{GEANT:2003}, which accounts for material properties, energy deposition, and signal formation processes~\cite{JINST:P07059}. The resulting detector signals were reconstructed with the same algorithms employed for the measured UMD data within the official \Offline software framework~\cite{Argiro:2007} of the Pierre Auger Observatory. This guarantees that simulated events are subject to the same reconstruction and selection procedures that would be applied to experimental data, providing a consistent and unbiased basis for validating the method.

\begin{figure}
\centering
\includegraphics[width=0.5\textwidth]{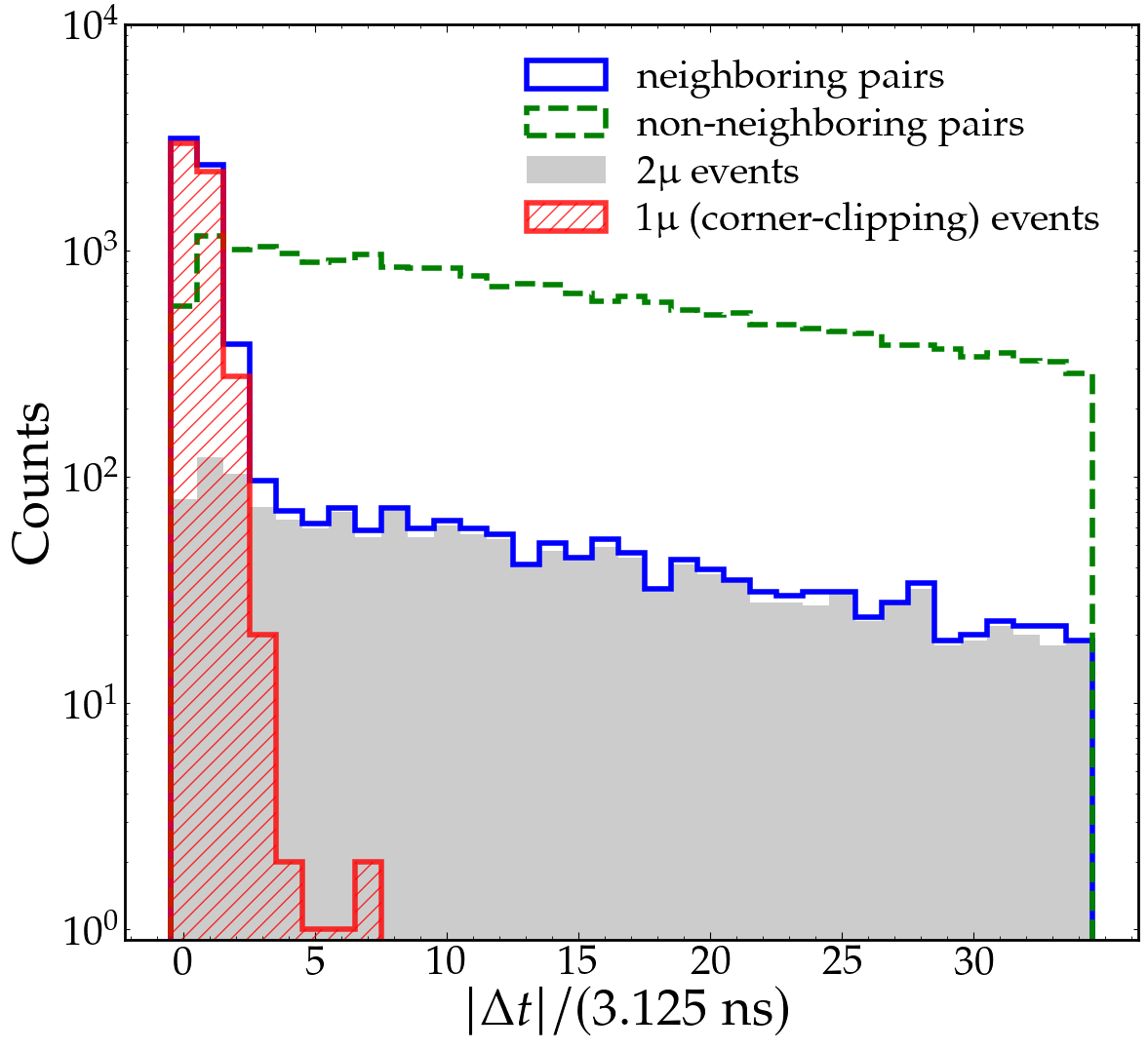}
\caption{Simulated distributions of the absolute time difference, $\lvert \Delta t \rvert$, for UMD module halves with exactly two activated strips ($\tilde{k}=2$). The distribution for non-neighboring pairs is shown by the green dashed line, while the distribution for neighboring pairs is represented by the blue solid line. Using Monte-Carlo truth, the neighboring-pair sample is decomposed into its two components: genuine single-muon corner-clipping events (red hatched area) and background two-muon events (grey filled area).}
\label{fig:sim_dt}
\end{figure}

We select simulated events in which a detector half exhibits exactly two activated strips ($\tilde{k} = 2$). We further require that the binary trace of each activated strip contains exclusively a single-muon pattern. Events in which either strip exhibits multiple patterns or complex structures indicative of pile-up are rejected. For the selected events, we construct the distribution of the absolute time difference, $\lvert \Delta t \rvert$, between the two signals. \Cref{fig:sim_dt} illustrates the resulting distributions. The distribution for non-neighboring pairs (green dashed line) is broad, consistent with the expectation for signals generated by two independent muons whose arrival times reflect the temporal spread of the shower front. In contrast, the distribution for neighboring pairs (blue solid line) exhibits a similar broad component but is distinguished by a sharp peak at small time differences $(\lvert \Delta t \rvert / \qty{3.125}{\ns} \lesssim 5)$. Utilizing the Monte-Carlo truth information, the neighboring-pair sample is decomposed into its constituent parts, which are genuine single-muon corner-clipping events (red hatched area) and background events originating from two independent muons (grey filled area). This decomposition confirms that the prominent peak observed at small $\lvert \Delta t \rvert$ is almost exclusively composed of corner-clipping events.

\begin{figure}
\centering
\includegraphics[width=0.48\textwidth]{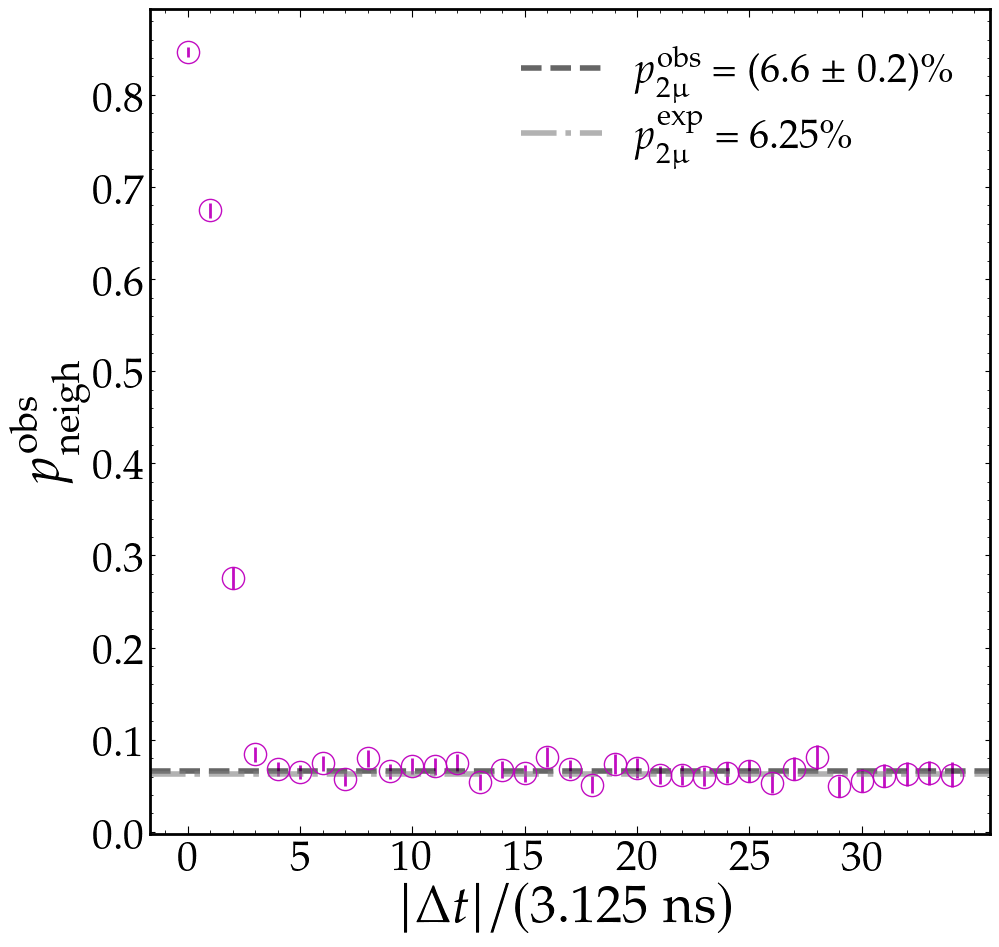}
\caption{The observed fraction of neighboring pairs, $p_{\text{neigh}}^{\text{obs}}$, in simulated events as a function of the absolute time difference, $\lvert \Delta t \rvert$. For large time differences $(\lvert \Delta t \rvert / \qty{3.125}{\ns} > 5)$, the fraction (open circles) reaches a constant asymptotic value. The average in this region yields $p_{2\upmu}^{\text{obs}} = \qty{6.6 \pm 0.2}{\percent}$ (dashed line), consistent with the theoretically expected combinatorial probability $p_{2\upmu}^{\text{exp}} = \qty{6.25}{\percent}$ (dash-dotted line).}
\label{fig:sim_frac}
\end{figure}

Next, we assess the performance of the background estimation procedure. \Cref{fig:sim_frac} displays the observed fraction of neighboring pairs, $p_{\text{neigh}}^{\text{obs}}$, as a function of the absolute time difference, $\lvert \Delta t \rvert$. Following the criteria established in \cref{sec:method}, the threshold at which the fraction reaches its plateau is identified as $\lvert \Delta t \rvert_{\text{cut}} / \qty{3.125}{\ns} = 5$. In the resulting control region, $\lvert \Delta t \rvert > \lvert \Delta t \rvert_{\text{cut}}$, the observed fraction (magenta circles) reaches a constant asymptotic value. The average in this regime yields the measured background probability $p_{2\upmu}^{\text{obs}} = \qty{6.6 \pm 0.2}{\percent}$ (dashed line), which is consistent with the theoretically expected combinatorial probability $p_{2\upmu}^{\text{exp}} = \qty{6.25}{\percent}$ (dash-dotted line). This agreement validates the identification of the large-$\lvert \Delta t \rvert$ region as a control sample dominated by random two-muon coincidences and effectively free of corner-clipping contamination. Conversely, as $\lvert \Delta t \rvert$ decreases, the fraction $p_{\text{neigh}}^{\text{obs}}$ increases sharply, indicating the onset and dominance of the corner-clipping signal at small time differences.

\begin{figure}
\centering
\includegraphics[width=0.5\textwidth]{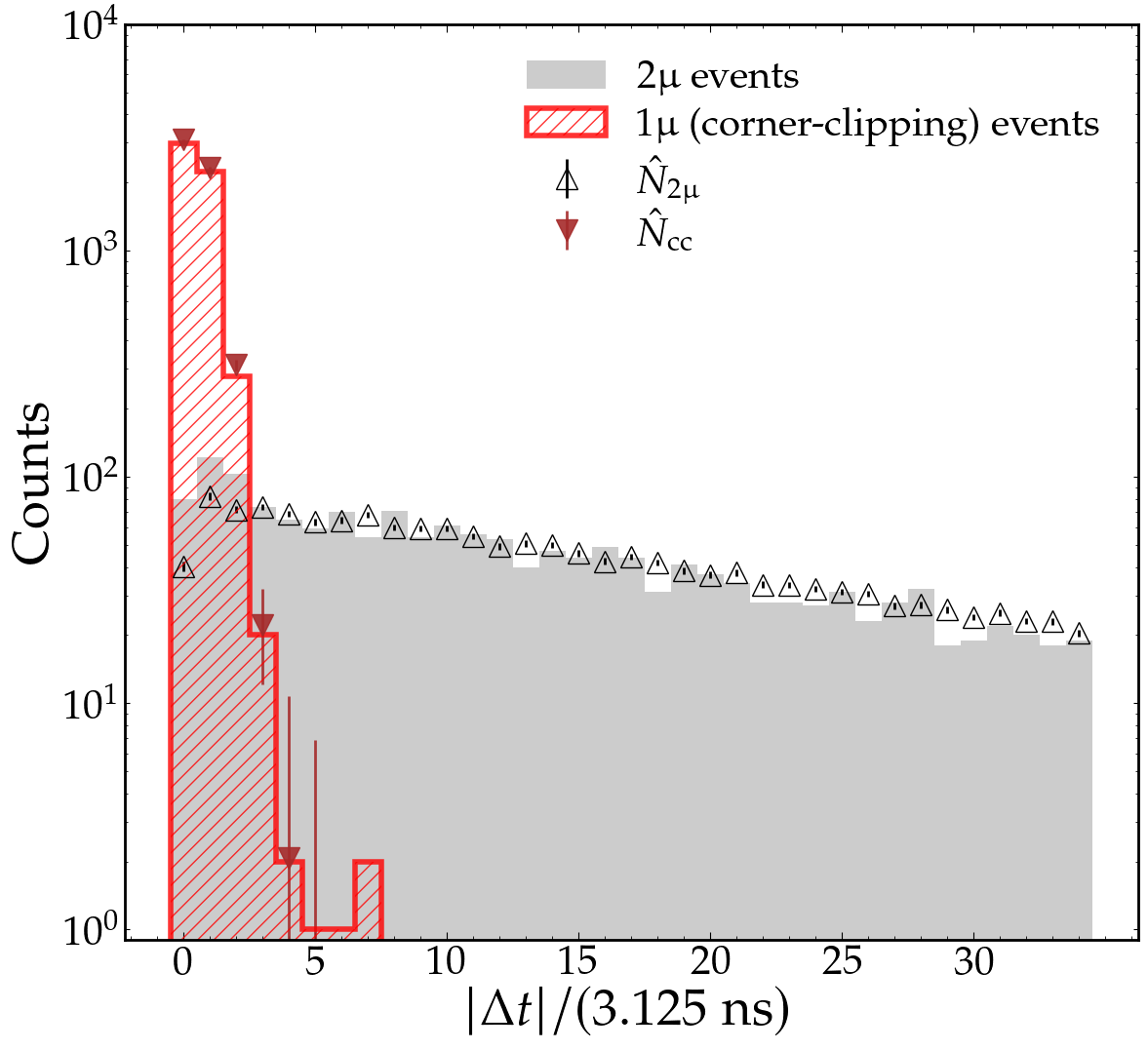}
\caption{Estimated numbers of corner-clipping events ($\hat{N}_{\text{cc}}$, solid inverted triangles) and two-muon background events ($\hat{N}_{2\upmu}$, open triangles) as a function of the absolute time difference, $\lvert \Delta t \rvert$, obtained using the data-driven method. For comparison, the true Monte-Carlo distributions of single-muon corner-clipping events (red hatched histogram) and two-muon events (grey filled histogram) are also shown. The close correspondence between the estimated and simulated distributions demonstrates the accuracy and robustness of the method in separating the two components.}
\label{fig:sim_dt_with_estimates}
\end{figure}

Using the background probability $p_{2\upmu}^{\text{obs}}$ obtained from the large-$\lvert \Delta t \rvert$ region, \cref{eq:Ncc_final2} and \cref{eq:N2mu_final2} were applied to the simulated data set to estimate, for each $\lvert \Delta t \rvert$ bin, the numbers of corner-clipping events ($\hat{N}_{\text{cc}}$) and two-muon background events ($\hat{N}_{2\upmu}$). The associated statistical uncertainties were evaluated using a bootstrap procedure~\cite{Efron:1979,Kulesa:2015}, which accounts for both statistical fluctuations and sample variability in the data-driven estimation. The resulting distributions are shown in \cref{fig:sim_dt_with_estimates}, where the estimates (triangular markers) are compared with the true Monte-Carlo values (filled histograms, previously shown in \cref{fig:sim_dt}). The excellent agreement between the data-driven estimates and the simulation truth demonstrates that the method accurately disentangles the corner-clipping signal from the two-muon background and reliably reconstructs both underlying components.

\begin{figure}
\centering
\includegraphics[width=0.48\textwidth]{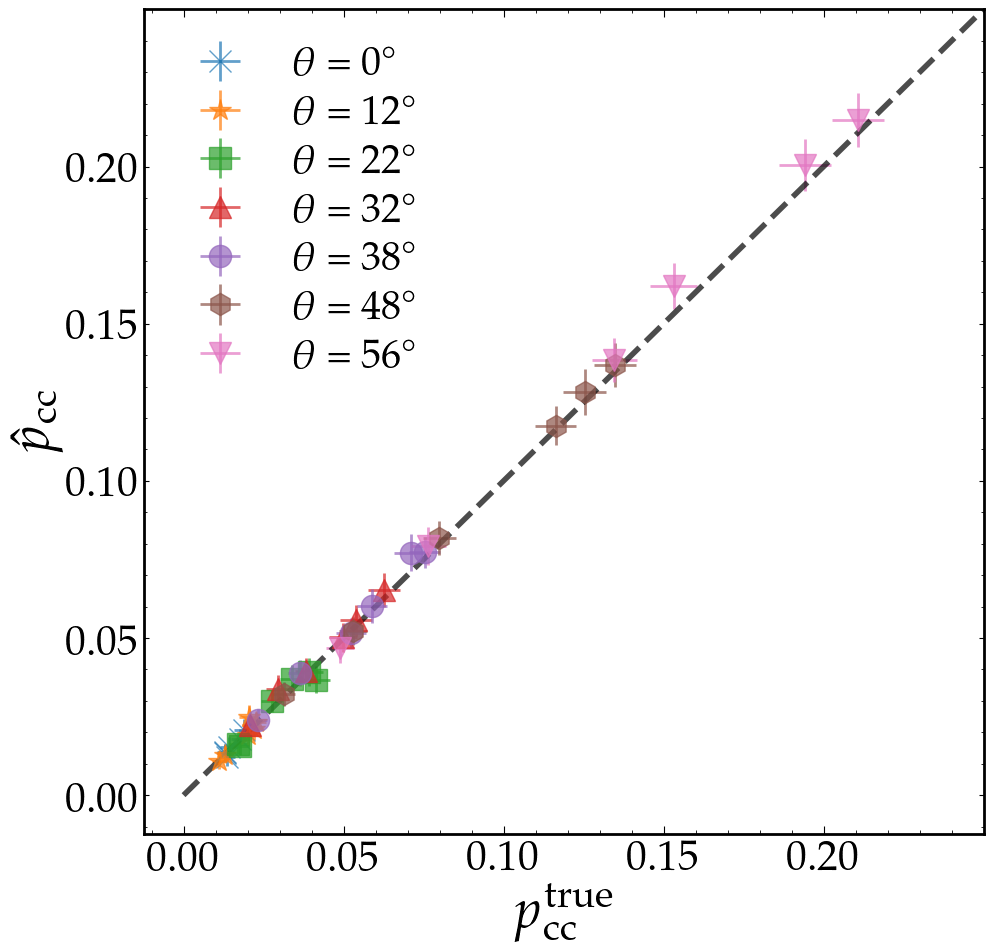}
\caption{The estimated corner-clipping probability, $\hat{p}_{\text{cc}}$, obtained with the proposed data-driven method, is plotted against the true probability, $p_{\text{cc}}^{\text{true}}$, from Monte-Carlo simulations. Each point represents a distinct bin in zenith and azimuth angle. The close alignment of the points with the identity line (dashed) demonstrates that the estimator accurately reproduces the true probability.}
\label{fig:pcc_validation}
\end{figure}

The final validation step compares the data-driven estimator $\hat{p}_{\text{cc}}$ (\cref{eq:pcc_est}) with the true probability $p_{\text{cc}}^{\text{true}}$ obtained directly from the simulation ground truth. The uncertainties on $\hat{p}_{\text{cc}}$ were also estimated using bootstrap resampling, while for the Monte-Carlo reference values, $p_{\text{cc}}^{\text{true}}$, the uncertainties were derived assuming binomial statistics as
\[
\sigma_{p_{\text{cc}}^{\text{true}}} = \sqrt{ \frac{ p_{\text{cc}}^{\text{true}} (1 - p_{\text{cc}}^{\text{true}}) }{ N_{1\upmu}^{\text{true}} } },
\]
where $N_{1\upmu}^{\text{true}}$ denotes the true total number of single-muon incidences on a sub-unit in each bin. \Cref{fig:pcc_validation} shows the comparison across all simulated bins of zenith and azimuth angle. The data points align closely with the identity line, confirming the accuracy of the estimator throughout the entire angular range. A quantitative assessment based on a linear fit of $\hat{p}_{\text{cc}}$ versus $p_{\text{cc}}^{\text{true}}$ yields a slope of $\num{1.03 \pm 0.02}$ and an intercept of $\num{-0.2 \pm 0.9 e-3}$, both consistent with unity and zero, respectively. The fit exhibits an excellent goodness-of-fit, with $\chi^2/n_{\text{df}} = \num{8.44} / \num{40}$, confirming that the estimator is unbiased and reproduces the true corner-clipping probability within statistical uncertainties. In particular, the corner-clipping probability reaches values up to ${\sim}\qty{20}{\percent}$ for showers with zenith angles near $\ang{56}$ and trajectories transverse to the scintillator strip orientation, representing a significant correction for muon-density reconstructions. The accurate recovery of $\hat{p}_{\text{cc}}$ across the full angular range confirms that the data-driven method performs reliably under these conditions.

\begin{figure}
\centering
\includegraphics[width=0.48\textwidth]{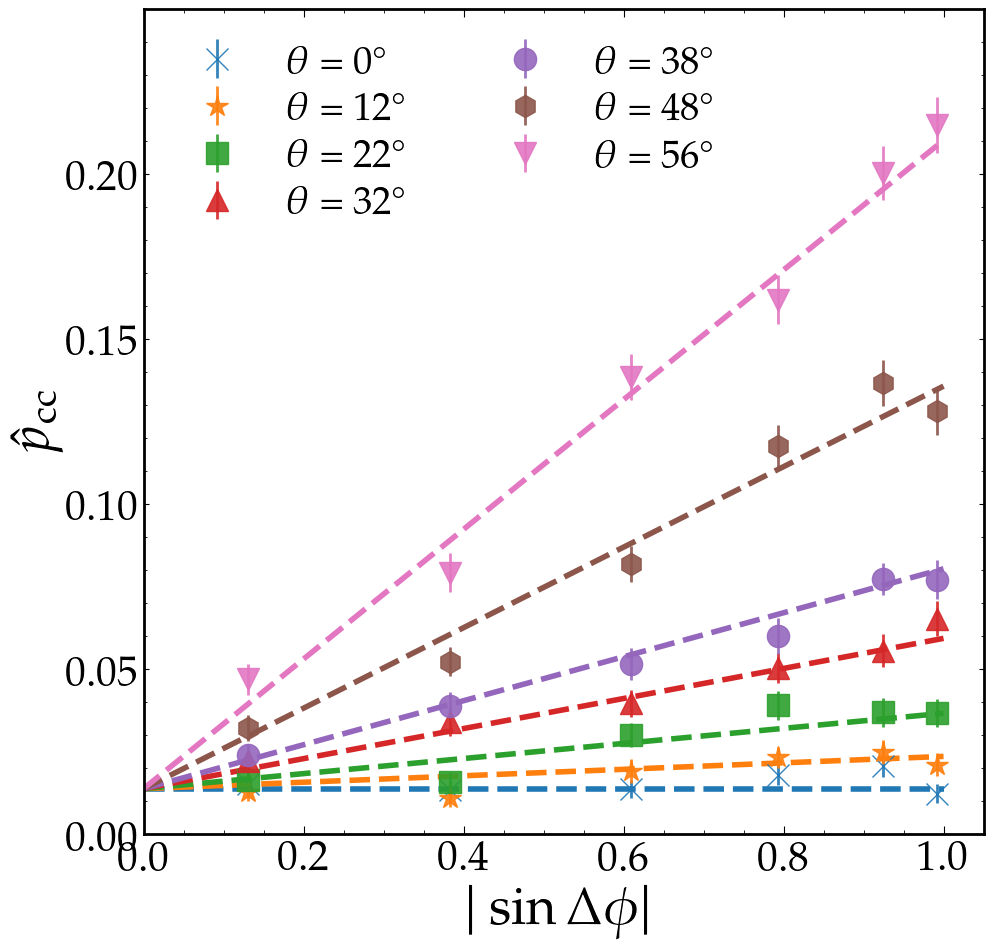}
\hfill
\includegraphics[width=0.48\textwidth]{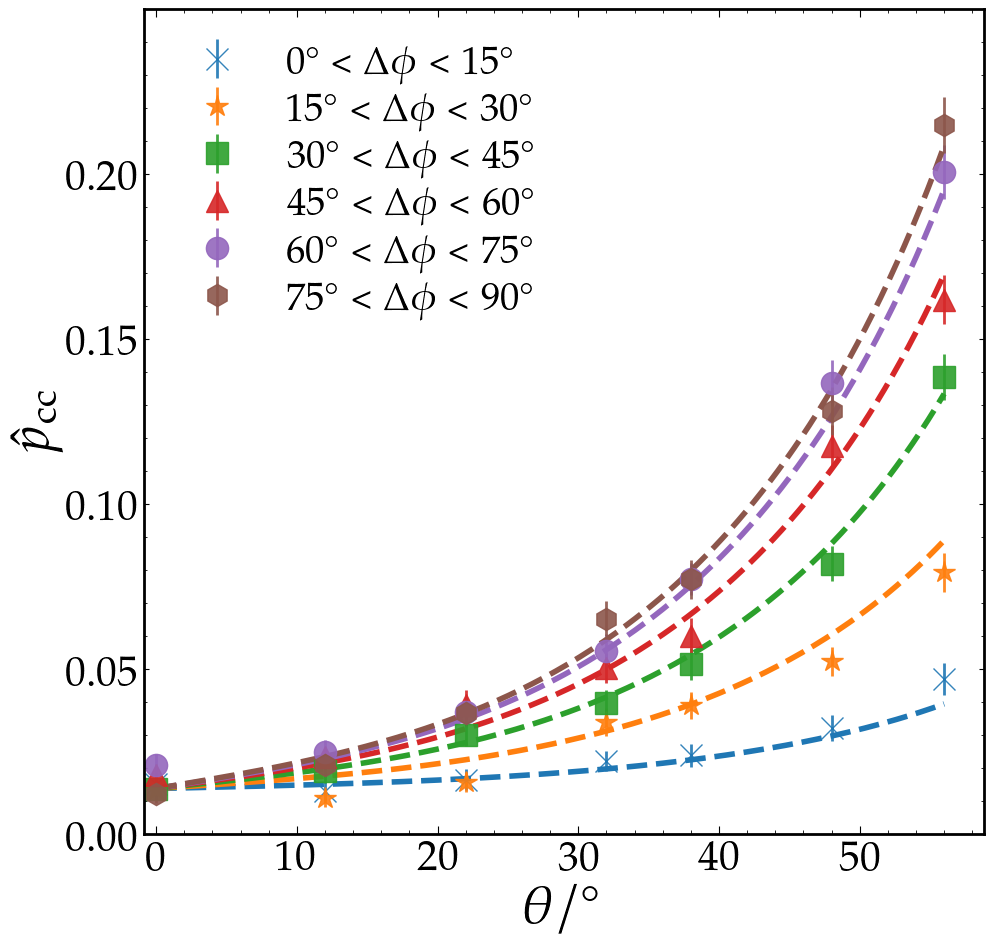}
\caption{
Estimated single-muon corner-clipping probability, $\hat{p}_{\text{cc}}$, for UMD simulated data as a function of $\lvert\sin(\Delta\phi)\rvert$ for seven zenith-angle values (top) and as a function of $\theta$ for six $\Delta\phi$ bins (bottom). The dashed lines correspond to the global fit using the parameterization defined in Eq.~\eqref{eq:pcc_param_geom}.}
\label{fig:pcc_simulated_data}
\end{figure}

\begin{figure}
\centering
\includegraphics[width=0.48\textwidth]{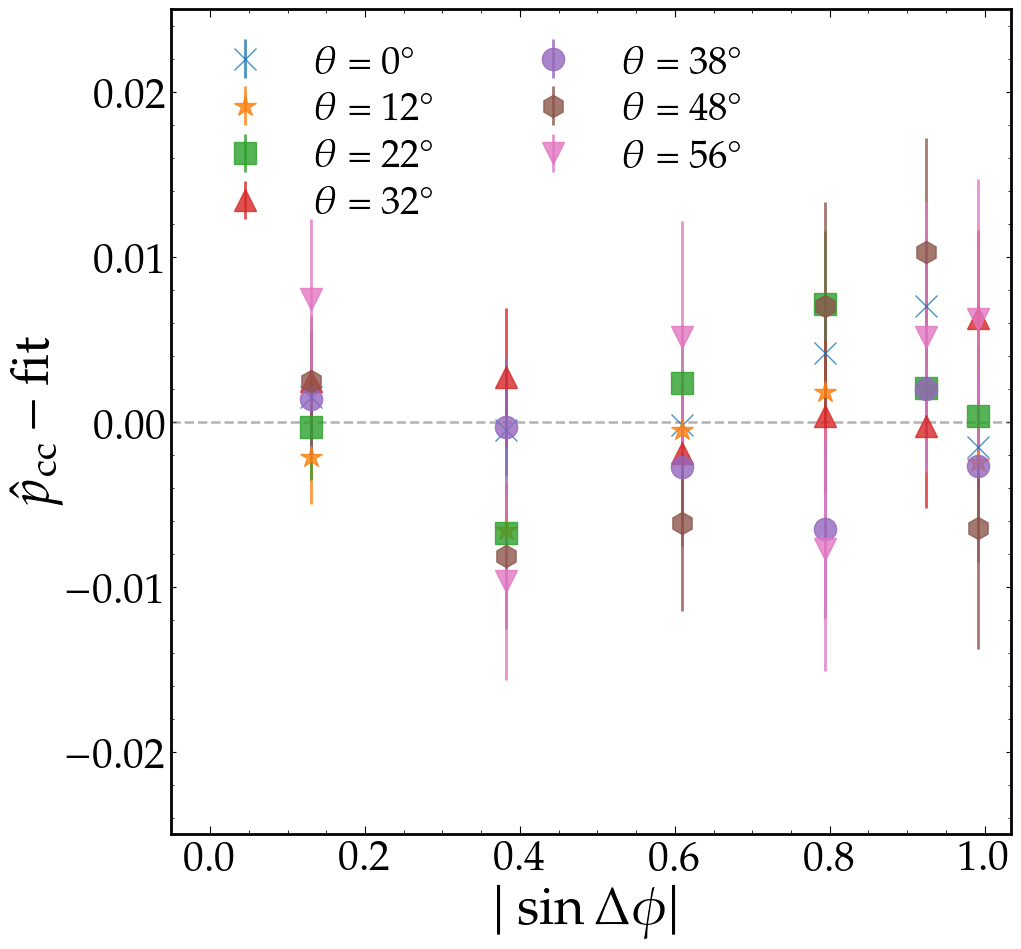}
\caption{Residuals of the global fit of the corner-clipping probability model to UMD simulated data, defined as the difference between the measured corner-clipping probability and the best-fit model from \cref{eq:pcc_param_geom}, shown as a function of $\lvert \sin(\Delta\phi) \rvert$ for different zenith angle bins. Residuals fluctuate around zero with absolute values below ${\sim} \num{0.01}$, indicating the high accuracy of the parameterization.}
\label{fig:pcc_fit_residuals_sims}
\end{figure}

The complete estimation procedure is applied to the simulated dataset to determine the corner-clipping probability, $\hat{p}_{\text{cc}}$.  The analysis is performed using six equally spaced azimuthal bins in the range $\ang{0} \le \Delta\phi \le \ang{90}$ and the seven discrete zenith angles available in the simulation library. As expected from geometric considerations, the results displayed in \cref{fig:pcc_simulated_data} show that $\hat{p}_{\text{cc}}$ increases with the zenith angle and, for a fixed $\theta$, rises as the particle trajectory becomes increasingly transverse to the detector strips.

To provide a functional form, we adopt the model derived in \cref{sec:model}, which incorporates both the geometric expectation for an ideal detector and corrections arising from real-detector effects. The parameterization is written as
\begin{equation}
    \hat{p}_{\text{cc}}(\theta,\Delta\phi) 
    = \big(A \, \tan\theta + B \, \sin\theta \big) \, \lvert \sin(\Delta\phi) \rvert + C,
    \label{eq:pcc_param_geom}
\end{equation}
where $A$, $B$, and $C$ are free parameters determined from a global fit to the data.

A global least-squares minimization of the model in \cref{eq:pcc_param_geom} was performed using the full simulated dataset, simultaneously constraining the parameters across all bins in $\theta$ and $\Delta\phi$. This procedure yields the best-fit values $A = \num{0.25 \pm 0.01}$, $B = \num{-0.21 \pm 0.01}$, and $C = \num{0.0138 \pm 0.0009}$. As anticipated in \cref{sec:model}, the numerical similarity between the trigonometric terms results in a strong anti-correlation between $A$ and $B$, with a Pearson correlation coefficient of $\rho_{AB} = \num{-0.97}$. This anti-correlation is a structural feature of the chosen basis functions and is enhanced when the angular coverage is concentrated at moderate zenith angles. Breaking this near-degeneracy requires angular coverage extending to sufficiently large zenith angles, where the two basis functions diverge: in the present simulation library, four of the seven sampled $\theta$ values lie in the region $\theta > \ang{30}$, providing the necessary lever arm. Despite this anti-correlation, the resulting curves reproduce the simulated angular dependence with excellent accuracy, as shown in \cref{fig:pcc_simulated_data}. The goodness-of-fit is confirmed by a $\chi^2/n_{\text{df}} = \num{44.3} / \num{39}$, and the residuals presented in \cref{fig:pcc_fit_residuals_sims} show no systematic structure, fluctuating randomly within a band of $\pm \num{0.01}$. These results demonstrate that the model successfully captures the full angular dependence of the corner-clipping probability, yielding a reliable and physically grounded parameterization.

To quantitatively assess the necessity of including both the geometric ($A$) and threshold-related ($B$) terms despite their high correlation, we performed reduced fits in which one of the two angular terms is suppressed. Setting $A = 0$ yields $\chi^2/n_{\text{df}} = \num{713.3}/\num{40}$, while setting $B = 0$ yields $\chi^2/n_{\text{df}} = \num{261.6}/\num{40}$, both dramatically worse than the full model. Following a standard likelihood-ratio test, the resulting changes in $\chi^2$ ($\Delta\chi^2 \approx \num{669}$ and $\Delta\chi^2 \approx \num{217.3}$, each for a single degree of freedom) rule out the single-term models at a significance far exceeding $5\sigma$. This confirms that both angular contributions are individually required and that the strong anti-correlation reflects the partial numerical overlap of two physically distinct terms rather than a genuine redundancy. That the fit successfully recovers the nominal geometric values under the idealized simulation conditions, as discussed below, provides direct confirmation that the extended angular coverage is sufficient to resolve the parameter degeneracy.

The fitted parameters align closely with expectations. The coefficient $A = \num{0.25 \pm 0.01}$ corresponds to the expected ratio $h/w$ for an ideal rectangular geometry. Using the ideal strip dimensions ($h = \qty{10.0}{\mm}$, $w = \qty{41.0}{\mm}$) gives $h/w = \num{0.244}$, in excellent agreement. The negative coefficient $B = \num{-0.21 \pm 0.01}$ reflects the corner-clipping efficiency reduction associated with the minimum detectable path length. From $B = -2d/w$, we infer a threshold path length $d = \qty{4.3 \pm 0.2}{\mm}$, which is consistent with the energy deposition required to exceed the discriminator threshold in the simulation framework. Finally, the small offset $C = \num{0.0138 \pm 0.0008}$ captures orientation-independent contributions. For the UMD, this term accounts not only for secondary knock-on electrons that occasionally activate neighboring strips, but also for the fact that $p_{\text{cc}}$ is evaluated as a function of the shower-axis direction rather than the individual muon trajectories. Since muons within an air shower exhibit an intrinsic angular dispersion, even vertical showers ($\theta = 0$) produce a non-zero residual corner-clipping probability. This effect, combined with the small but irreducible contribution from secondary particles, naturally explains the magnitude of the fitted $C$ parameter.

\section{Conclusions}
\label{sec:conclusions}

The corner-clipping of particles in segmented detectors introduces a direction-dependent overcounting bias that can affect precision particle counting. We have presented a data-driven method to determine the single-particle corner-clipping probability, $p_{\text{cc}}$, using timing information and the topology of activated detector segments. The method separates nearly simultaneous correlated activations from accidental coincidences by using the large-$\lvert \Delta t \rvert$ region and non-neighboring segment pairs as an intrinsic control sample.

The methodology was validated using detailed simulations of the Underground Muon Detector (UMD) of the Pierre Auger Observatory. Across the full range of simulated shower energies and geometries, the estimator accurately recovered the true single-particle corner-clipping probability. A linear fit of $\hat{p}_{\text{cc}}$ versus $p_{\text{cc}}^{\text{true}}$ yielded a slope of $\num{1.03 \pm 0.02}$ and an intercept of $\num{-0.2 \pm 0.9 e-3}$, both statistically consistent with the identity relation. Notably, the method performs reliably at $\theta = \ang{56}$, where the corner-clipping probability reaches values up to ${\sim}\qty{20}{\percent}$ for trajectories nearly transverse to the scintillator strips. This demonstrates the potential of the approach to extend the physics reach of the detector into angular regimes where simulation-based corrections were previously considered uncertain.

To parameterize the results, we introduced a compact analytical model incorporating the geometric scaling of rectangular segments, the efficiency reduction due to a finite detectable path length, and an orientation-independent term. Despite the strong structural anti-correlation between the geometric and threshold-related coefficients ($\rho_{AB} \approx -0.97$), likelihood-ratio tests confirm that both terms are individually required to describe the results. The best-fit coefficients align closely with known detector properties: the parameter $A$ recovers the nominal geometric ratio $h/w = 0.244$, while the parameter $B$ implies a threshold path length $d = \qty{4.3 \pm 0.2}{\mm}$, consistent with the energy deposition required to exceed the discriminator threshold. The constant term $C$ successfully captures residual contributions from secondary particles and the intrinsic angular dispersion of muons relative to the shower axis.

In experimental applications, the measured function $p_{\text{cc}}(\theta,\Delta\phi)$ can be incorporated directly into reconstruction algorithms to mitigate the overcounting bias and improve the determination of the muonic component of extensive air showers. More generally, the estimation procedure and analytical parameterization are not specific to the UMD and can be applied to other segmented detectors with sufficient timing resolution and a well-defined adjacency structure.

\begin{acknowledgement}
This work received support from the Comisión Nacional de Energía Atómica (CNEA), the Consejo Nacional de Investigaciones Científicas y Técnicas (CONICET), and the Universidad Nacional de San Martín (UNSAM) in Argentina, as well as from the Karlsruhe Institute of Technology (KIT) in Germany. Partial funding was provided by CONICET under grant PIP 2023-2025 GI 11220220100586CO.
\end{acknowledgement}


\end{document}